\documentclass[aps,preprint,groupedaddress]{revtex4-2}

\usepackage{graphicx}

\newcommand{\lgdx}{\log_{10}}

\begin{document}


\title{Tidal disruptions of rotating stars by a supermassive black hole}


\author{Kazuki Kagaya}\thanks{Present affiliation: Terrabyte Co., Ltd.}
\author{Shin'ichirou Yoshida}
\email{yoshida@ea.c.u-tokyo.ac.jp}
\author{Ataru Tanikawa}
\affiliation{Department of Earth Science and Astronomy, Graduate School of Arts and Sciences, The University of Tokyo, \\Komaba 3-8-1, Meguro-ku, Tokyo 153-8902, Japan}


\date{\today}

\begin{abstract}
We study tidal disruption events of rotating stars by a supermassive black hole in a galactic nucleus
by using a smoothed-particle hydrodynamics (SPH) code.
We compare mass infall rates of tidal-disruption debris of a non-rotating and of a rotating star when they
come close to the supermassive black hole. 
Remarkably the mass distribution of debris bound to the black hole as a function of specific energy 
shows clear difference between rotating and non-rotating stars, even if the stellar rotation is far from the break-up limit.
The debris of a star whose initial spin is parallel to the orbital angular momentum has a mass distribution
which extends to lower energy than that of non-rotating star. The debris of a star with anti-parallel
spin has a larger energy compared with a non-rotating counterpart. As a result, debris from a star with anti-parallel spin is bound more loosely  to the black hole and the mass-infall rate rises later in time, while that of a star with a parallel spin is tightly
bound and falls back to the black hole earlier. The different rising timescales of mass-infall 
rate may affect the early phase
of flares due to the tidal disruptions.

In the Appendix we study the disruptions by using a uniform-density ellipsoid model
which approximately takes into account the effect of strong gravity of the black hole.
We find the mass infall rate reaches its maximum earlier for strong gravity cases
because the debris is trapped in a deeper potential well of the black hole.
\end{abstract}


\maketitle

\section{Introduction}
Observational evidences have been accumulating that most of large
galaxies have a supermassive black hole (SMBH) of mass $M_{\rm BH}$
in the range of $10^6\le M_{\rm BH}/M_\odot\le 10^{10
}$ at their centers. Some of these SMBHs
show continuous cataclysmic activities known as active galactic
nuclei (AGN). AGN are thought to occur in a environment where gas is supplied to 
SMBH in sufficiently high rates \citep{Rees1984ARAA}.
In an opposite environment with little gas supply to SMBH, transient
activities are still possible with a star or gas cloud occasionally
accreting to SMBH. A tidal disruption event (TDE) occurs when the 
star or the gas cloud passes close enough to be disrupted by the tidal
force of SMBH\citep{Hills1975Natur, FrankRees1976, KatoHoshi1978}. 
Around the periastron of the object's orbit, the tidal force of SMBH 
exceeds the self-gravity of the object and tears it apart. A part
of disrupted debris then accretes onto SMBH and radiates mainly
in ultra-violet to soft X-ray band. 

Although details of hydrodynamic and radiation processes of the disruptions are not fully known, a simple model predicts an important aspect of the light curves of the TDEs
\citep{Lacy1982, Rees1988Natur, Phinney1989, EvansKochanek1989}. A star passing by a SMBH whose
periastron distance is smaller than the tidal radius is instantaneously disrupted.
The fluid elements of the disrupted object follow Keplerian orbits around the black hole
which are determined by the total specific energy $E$. The orbital period $T$ is related to
$E$ by 
\begin{equation}
	E = -\frac{1}{2}\left(\frac{2\pi G M_{\rm BH}}{T}\right)^\frac{2}{3},
	\label{eq: energy of fluid element}
\end{equation}
when $E$ is negative (i.e., the fluid element is bound to the black hole). Differentiating
Eq.(\ref{eq: energy of fluid element}) with respect to $T$ and multiplying it by the differential mass distribution of
the fluid elements as a function of $E$, $dM/dE$, we have
\begin{equation}
	\frac{dM}{dT} = \frac{dM}{dE}\frac{dE}{dT}=\frac{(2\pi G M_{\rm BH})^{\frac{2}{3}}}{3}\frac{dM}{dE} T^{-\frac{5}{3}}.
	\label{eq: mass infall rate}
\end{equation}
The fluid elements return to the point close to the periastron after the time interval of $T$. They collide each other to form a disk, in which the gas accretes in a time scale much shorter than the orbital period.
Thus we expect the fall-back rate $dM/dT$ to be proportional to the luminosity of the event.
If the mass distribution $dM/dE$ is flat, Eq.(\ref{eq: mass infall rate})  shows the luminosity
changes in time as $\sim T^{-5/3}$. The numerical simulations actually shows that $dM/dE$
have a flat profile around $E=0$ (see e.g. \cite{EvansKochanek1989}). Thus the time evolution
of luminosity for the later stage
of the TDEs is simply represented by the $T^{-5/3}$-law. This
is observed in X-ray \citep{Komossa2012} and in optical band \citep{Hung_etal2017}.

The ratio of the tidal radius $R_t$ to the Schwarzschild radius $R_s$ of the central black hole is
\begin{equation}
	\frac{R_t}{R_s} = 24\left(\frac{M_{\rm BH}}{10^6M_\odot}\right)^{-\frac{2}{3}}\left(\frac{R_\star}{R_\odot}\right)
	\left(\frac{M_\star}{M_\odot}\right)^{-\frac{1}{3}},
\end{equation}
by which we see a TDE of solar type star occurs in strong gravity region of the spacetime for 
$M_{\rm BH}\sim 10^6 M_\odot$. Thus general relativistic effects have been taken into account in recent modeling of TDEs \citep{Franchini2016MNRAS,Hayasaki2016MNRAS,Tejeda2017MNRAS}.
One of the main issues is observational consequence of such strong gravity effect as dragging of
inertial frame \citep{StoneLoeb2012PhRvL}.

In the early phase of the TDEs, the light curve may not necessarily follow the
simple model above. In fact, \cite{LodatoKingPringle2009} examined the effect of the internal structure 
of the incident star on the TDEs (see also \cite{GuillochonRamirez-Ruiz2013}).
They show that the higher concentration of density in the star results in the slower rise of the light
curve. Further observational signatures of more realistic stellar structures are examined 
by \cite{2018arXiv180103497G}, in which a time evolution of chemical composition in 
tidal disruption flare is discussed.

Structure of the incident star is also affected by stellar
spin. Centrifugal force comes into the hydrostatic balance between
self-gravity and pressure gradient. Observationally non-negligible
fractions of early type stars have large angular momentum so that the
stellar structure significantly deviates from spherical symmetry
\citep{vanBelle2012}.  Although the event rate of tidal disruptions
for rotating stars is not known,
\cite{Kochanek2016} investigates the event rate of TDE's as a function of
black hole mass as well as mass of different type of stars for different histories of
star formations. It is shown the rate for stars more massive than
$1M_\odot$ is rather small and the TDE events are
dominated by the stars with the smaller masses. Considering the
observed stellar surface velocity tends to be slow for late-type stars
\citep{Tassoul1978, Palacios2013}, possibly due to angular momentum
loss by stellar wind or magnetic breaking, there may not be much chance of
observing TDE of rapidly rotating early type stars.  Still it may be
useful to study the possible outcome of the stellar rotation in the
TDEs. Although {\tt MOSFiT} \citep{2018ApJS..236....6G} can
  generate light curves for TDEs with wide parameter ranges, it does
  not take into accout the stellar rotation. TDE light curves would
  not be determined only by the fall-back rate
  \citep{2015ApJ...804...85S,2015ApJ...806..164P}, however the
  fall-back rate is still an important property for TDEs. In this
paper we examine the effect of stellar rotation on the mass infall
rate after the tidal disruption by using a non-linear hydrodynamic
simulation as well as a simple ellipsoidal model of a star.  As in
precedent works, we evaluate the
energy of the tidally disrupted debris as a proxy of the mass infall rate 
by assuming its Keplerian motion around SMBH.

The paper is organized as follows. In Sec.\ref{sec: SPH} we briefly
introduce our setup of hydrodynamic simulations. In Sec.\ref{sec: Results}
we present our main results. Finally in Sec.\ref{sec: Summary} we summarize
our results and make some remarks. In Appendix \ref{sec:affine model}
we study effects of strong gravity of supermassive black hole on the tidally
disrupted debris by using
a simple homogenous ellipsoid model of stars.

\section{\label{sec: SPH}Numerical Simulations}

	\subsection{Numerical simulation code}
We perform three-dimensional smooth particle hydrodynamic (SPH) simulations by using
numerical code developed in \cite{Tanikawa2017} (see
  also \cite{2018MNRAS.475L..67T,2018ApJ...858...26T}).  
  Cartesian coordinate is used in the code. 
  Wendland C2
kernel for the SPH kernel interpolation \citep{Wendland1995,
  Dehnen-Aly2012} is adopted.  The number of neighbor particles of a
given particle is about 120 (3D).  The artificial viscosity proposed
by \cite{Monaghan1997} is used. The viscosity from shear motion is
suppressed by the Balsara switch \citep{Balsara1995}. Self gravity
among SPH particles is computed with adaptive gravitational softening
\citep{Price-Monaghan2007}. The SPH and self-gravity calculations are
optimized on distributed-memory systems, using FDPS \citep{Iwasawa2015, Iwasawa2016}
 and explicit AVX instructions
\citep{Tanikawa2012,Tanikawa2013}.

  	\subsection{Rotating stars}
\begin{figure}
\includegraphics{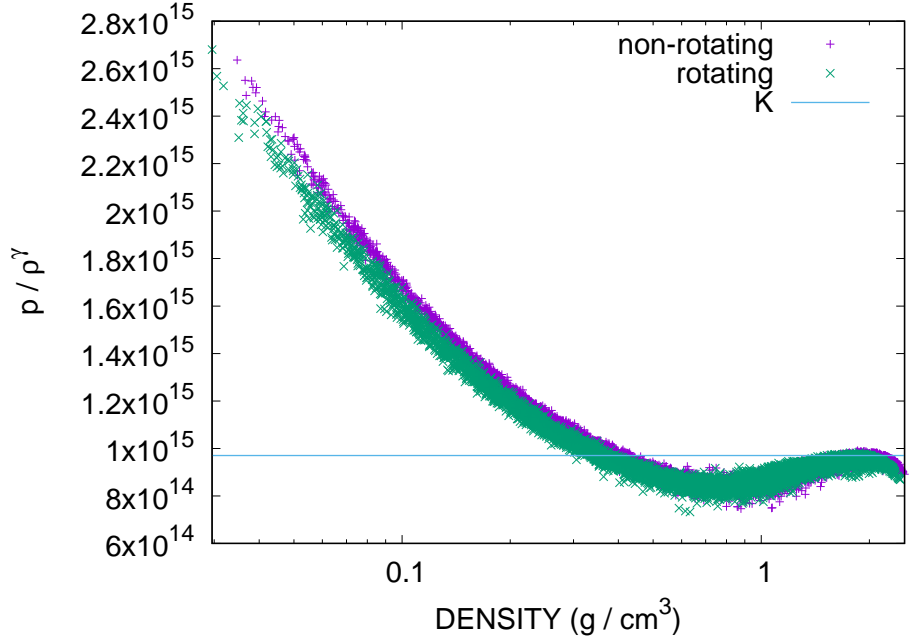}
\caption{Comparison of $p/\rho^\gamma$ as a function of density $\rho$ after the SPH relaxation process.
The original models of stationary star have polytropic index $N=1.5$.
The constant $K$ is the polytropic constant of the equilibrium model which is used to construct the
initial data by the SPH relaxation.}
\label{fig: entropy}
\end{figure}

We compute uniformly rotating stellar models by using a numerical code \citep{Yoshida_Eriguchi1995} based on
Hachisu's self-consistent method \citep{HSCF}. The code computes stationary and axisymmetric
rotating stellar equilibria on the spherical polar-coordinate grids. The models need to be projected
onto the Cartesian coordinate grids to prepare the initial stellar models for the numerical simulations.
Although the projection is mathematically straightforward, the initial models produced on the finite grids require
relaxation runs in the SPH code before performing simulations of tidal disruptions. A projected
model on the Cartesian grid is evolved by the SPH code until we have a stationary state. Due to the
numerical viscosity in the code, entropy is produced in the model during the relaxation run which
lasts two to three dynamical timescales of the star. 
In Fig.\ref{fig: entropy} we plot the values of $p/\rho^\gamma$ as a function of $\rho$, which is a proxy
of entropy distribution. 
For each fluid element the value is computed after the relaxation procedure.
For the polytropic fluid used to construct the equilibrium model, the value would be constant.
The deviation of it from being constant results from the fact that the code suffers
from numerical dissipations. Nevertheless the fluids composing the relaxed models 
for non-rotating and rotating cases are 
identified when the profiles of $p/\rho^\gamma$ overlap as in Fig.\ref{fig: entropy}.

In our numerical simulations we focus on stellar models with $\gamma=5/3$ and the mass of $4.0M_\odot$.
The stellar rotation is parametrized by the so-called 'T/W' value, which is the ratio of kinetic energy
to the gravitational energy\citep{HSCF}. We choose the star to have a polar-to-equatorial axis ratio
of $0.95$, which corresponds to $T/W=8.3\times 10^{-3}$. The rotational velocity at the equator of the
stars amounts to $300~\rm{km}~\rm{s}^{-1}$. 

\section{Results\label{sec: Results}}

A black hole with mass $M_{\rm BH}=10^6M_\odot$ is placed at the coordinate origin. We assume the mass of stars is negligibly 
small compared to that of the black hole, thus we fix the position of the black hole. In this approximation the center of the 
mass of the star moves along a conic section. We choose the orbit of the center of the mass to be a parabola
whose periastron distance is half the tidal radius $R_t$. The initial distance of the star from the black hole is twice as far
as $R_t$. We performed simulations with stars whose rotational axis is perpendicular to the orbital plane. The model is
named as 'parallel spin' when the spin angular momentum of the star is parallel to that of the orbit. When the spin
angular momentum is parallel to the orbital one but with a different sign, it is called 'anti-parallel spin' case. We also performed
a simulation with an initially non-rotating star ('zero spin' case). 

\begin{figure}
\includegraphics{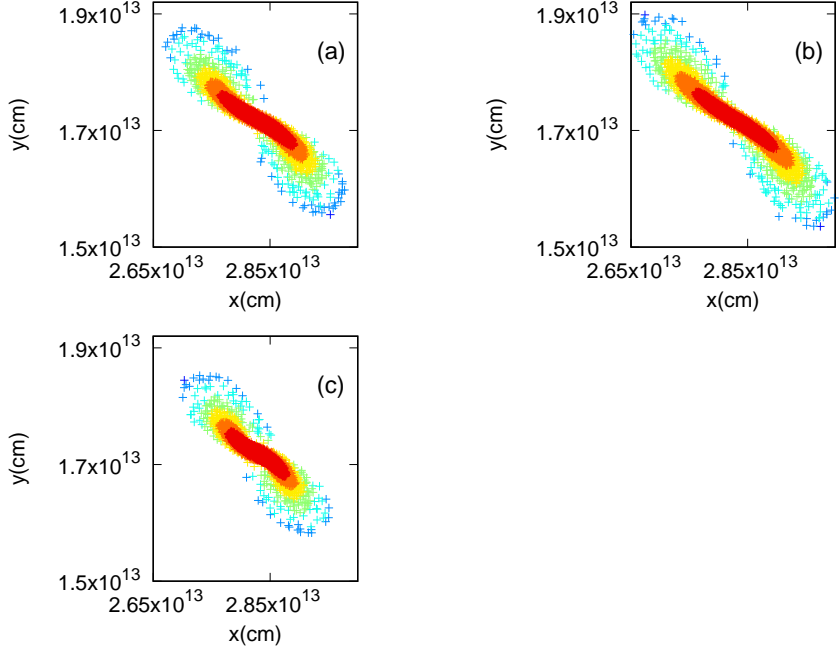}
\caption{Snapshot of SPH particle distributions projected 
onto the orbital plane. In each panel, the black hole is at the origin. The top left panel (a) is for the non-rotating stellar model. 
The top right (b) and the bottom (c) correspond to the stars with parallel and anti-parallel spins
respectively. Dimensionless parameter $T/W$ of initial stellar spin is $8.3\times 10^{-3}$.
The contour map corresponds to the mass density.  From the innermost  to the outermost
region, the corresponding density is as follows: $-1.5< \lgdx\rho< 0.0$, 
$-2.0< \lgdx\rho_0 < -1.5$,
$-2.5< \lgdx\rho_0 < -2.0$,
$-3.0< \lgdx\rho_0 < -2.5$,
$-3.5< \lgdx\rho_0 < -3.0$,
$-4.0< \lgdx\rho_0 < -3.5$ and 
$-4.5< \lgdx\rho_0 < -4.0$, where $\rho_0\equiv \rho/1{\rm g}~{\rm cm}^{-3}$.
}
\label{fig: snapshot}
\end{figure}
\begin{figure}
\includegraphics{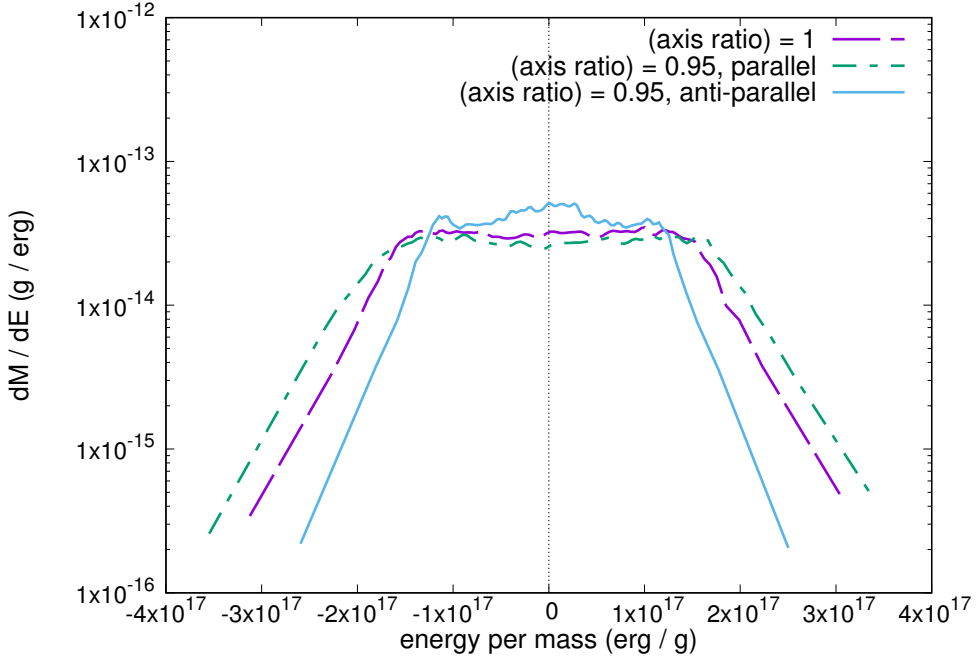}
\caption{Mass distribution of TDE debris as a function of mechanical energy of fluid element. 
Long-dashed line is for an initially
non-rotating star of $4M_\odot$. The dashed-dotted line is for a spinning star whose spin angular momentum 
is parallel to the orbital angular momentum while the solid line corresponds to the star whose spin angular momentum is anti-parallel to the orbital
angular momentum. The rotation of the model is parameterized by $T/W=8.3\times 10^{-3}$
or normalized rotational angular frequency of $\omega/\sqrt{\pi G\rho_c}=0.13$.
All the model is polytropic with $N=1.5$.}
\label{fig: dMdE}
\end{figure}

In Fig.\ref{fig: snapshot} the distributions of SPH particles after the stars passed the periastron are shown. 
The distribution of fluid elements are viewed from the direction perpendicular to the orbital plane. The orbital
angular momentum is pointing toward us. In this figure, particles off the orbital plane are projected onto the plane.
The snapshots are for the time lapse $t=2 t_0$, where $t_0$ is the time it takes for a star to reach its periastron. 
The figure shows tidally elongated structures of stellar fluid. A remarkable difference is that the distribution in the parallel
spin star is less compact than the zero spin case. 
For the anti-parallel spin model, on the other hand, the distribution is more compact than 
that of the zero spin case. The expansion and contraction of the particles distributions compared to the zero spin case result in the difference
in the mass distribution per specific energy as is seen in Fig.\ref{fig: dMdE}. Here differential mass of fluid per unit mechanical 
(kinetic plus gravitational) energy $dM/dE$ is plotted as a function of specific mechanical energy.  The fluid elements having
negative specific energy are regarded as being bound to the central black hole after the disruption. They accrete to the black hole
and flare up. The plateau around the zero energy corresponds to the fluid elements which are loosely bound to the black hole.
Because of their flat profile of $dM/dE$ and the long orbital period (see Eq.(\ref{eq: energy of fluid element})), 
they regulate the late-time canonical behaviour of the TDE events ("$T^{-5/3}$- law" as seen in Eq.(\ref{eq: mass infall rate})).
Beyond the edge of the plateau the distribution of $dM/dE$ gradually decreases as the specific energy decreases. Fluid
elements in this tail are deeply bound to the black hole and accrete faster after the disruption. They are expected to 
determine the early rises of the TDE flares. It is remarkable in this context that the parallel spin star has the most extended
low energy tail among the three cases, while the anti-parallel spin case has the shortest tail as seen in Fig.\ref{fig: dMdE}.
It reflects the fact that parallel spin case has the most extended debris by the tidal disruption and has more fluid
elements strongly bound to the potential of the black hole. The anti-parallel spin star, on the other hand, has
the most compact debris and the fluid elements is weakly bound to the black hole.

\begin{figure}
\includegraphics{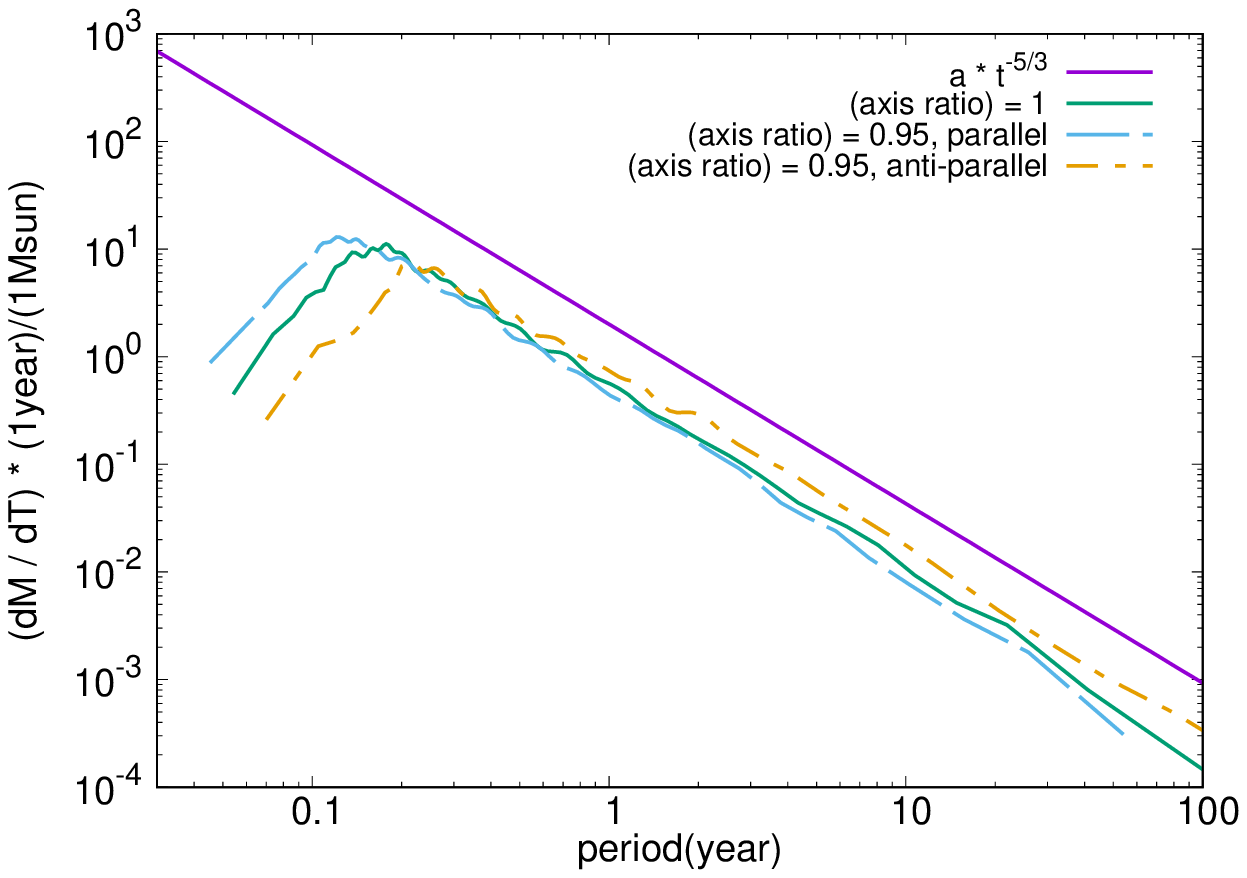}
\caption{Mass infall rate computed from $dM/dE$ profile in Fig.\ref{fig: dMdE}. 
Solid line is for a zero spin star, while dashed line is for a parallel spinning star 
and dashed-dotted is for a anti-parallel spinning star.
}
\label{fig: dMdT}
\end{figure}

We see a difference of mass infall rate for the three cases as a consequence of the difference in the debris 
distributions. As $dM/dE$ appears in the mass infall rate of Eq.(\ref{eq: mass infall rate}), we directly compare the rate for 
these cases (Fig.\ref{fig: dMdT}). 
As seen in the figure, the late time behaviour of mass infall rate scales as power law of elapsed time $t$
with the expected index $-5/3$. In the early phase the mass infall rate for the parallel spin case rises fastest and
the peak infall rate is the highest among three cases. The anti-parallel spin case is the slowest in rise
of mass infall rate and the lowest at its peak. This behavior is expected from Figs.\ref{fig: snapshot} and
\ref{fig: dMdE} since the parallel spin case has more fluid elements being trapped deep in the potential 
well of the black hole and the fluid elements of the anti-parallel spin case is more tightly bound to each other.

The physics behind the difference is rather simple. When we see the star in a comoving
(accelerating) frame with it, the stellar fluid suffers from inertial forces due to the
orbital motion. The centrifugal and Euler forces are canceled out, thus we are left with
the Coriolis force acting on the stellar fluid. If the star is spinning with the angular velocity
$\vec{\omega}$, the Coriolis force acting on the fluid at $\vec{r}$ (the position vector 
measured from the stellar center) is $2 (\vec{r}\times\vec{\omega})\times \vec{\Omega}$,
where $\vec{\Omega}$ is the instantaneous orbital angular velocity of the center of mass of the star.
The ratio of the Coriolis force to the centrifugal force from the stellar spin is thus
$2\Omega/\omega$. In our case the ratio is at least ${\cal O}[10]$. Moreover
the direction of the Coriolis force is such that it makes the star expand when the stellar
spin is parallel, while the force make the star contract when the spin is anti-parallel.

\section{Summary and remarks\label{sec: Summary}}
We studied influences of stellar spin on tidal disruption of a star by a supermassive
black hole. By using SPH simulations we compared the
tidally-disrupted debris of stellar model with the same mass but with different spins.
We found the distribution of debris, when the stellar spin is initially
anti-parallel to its orbital angular momentum, is more compact than that of a non-rotating star.
On the other hand the debris distribution is less compact for a star with parallel
spin. This leads to the difference in the initial rise of the mass-infall rate of debris.
It occurs earlier than the non-rotating case for a parallel spin star whose angular momentum
is aligned to the orbital one. It occurs later for an anti-parallel spin case. whose spin direction is opposite
to the orbital one. The difference comes
from the Coriolis force acting on the stellar fluid. The force tends to bind the fluid together
when the spin is anti-parallel to the orbital angular momentum, 
while it tends to tear it apart when the spin is parallel to the orbit. 
The force is at least an order of magnitude
larger than that of the centrifugal force by the stellar rotation itself.
It should be noted that the mechanism we see here 
may be also applied to totally different systems. For instance,  a planet or planetesimal may be
tidally disrupted at its close encounter to a star. A parallel spin planet to its orbit is more 
susceptible to tidal disruption than an anti-parallel spin one. A planet with an anti-parallel 
spin is more likely to survive in its close encounter to the central star.
Another case of interest is a tidal disruption of a dwarf satellite galaxy by a larger
one. The disruption is more likely to occur if the rotation of the dwarf is aligned to its orbital
angular momentum. Finally, a binary of stars or black holes orbiting around a massive
black hole may be bound tighter if the angular momentum of the binary system 
is anti-aligned to the orbital angular momentum of the binary around the massive black hole.
A breakup of a binary is more likely to occur if the binary motion is in the same direction as that
of the orbit of the binary around a massive black hole.

\appendix*
\section{Ellipsoidal (Affine) model\label{sec:affine model}}
In this appendix we investigate the difference in TDE mass infall rate arising from the stellar rotation
by using a simple ellipsoidal (Affine star) model \citep{CarterLuminet1985MNRAS.212...23C}. 
The model has been used to study tidal interactions of stars orbiting around supermassive black holes
as well as quasi-equilibrium configurations of close binary stars \citep{Lai1993ApJS...88..205L}.
We follow a simplified version of Affine star model formulated by \citet{UsamiFujimoto1997ApJ...487..489U}.
A star is approximated to be homogeneous, compressible and uniformly rotating while the other 
component (SMBH) is regarded as a point source of gravity. We assume the mass of
the star $m_\star$ is much smaller than that of the black hole $M_{\rm BH}$. 
An orbit of the star around the black hole is expressed by a conic section. We introduce
a Cartesian coordinate centred at the black hole whose z-axis is perpendicular to the orbital
plane of the star. Then the geometrical center of the star is expressed as $\vec{R} = (R(t)\cos\varphi(t),R(t)\sin\varphi(t),0)$.
The equilibrium figure of the star is assumed
to be ellipsoidal, which enables the profiles of pressure and scalar potential of self-gravity to be expressed
by quadratic forms of the Cartesian coordinate centred at the geometrical center of the star. Three principal axes
of the ellipsoidal star $a_1, a_2, a_3$ are functions of time. We denote $a_1$ as the semi-major axis
and $a_3$ is the principal axis perpendicular to the orbital plane. We have $a_1\ge a_2\ge a_3$. 
We consider the case in which the stellar spin angular momentum is perpendicular to the orbital
plane. The vorticity of the stellar fluid $\lambda(t)$, thus, has only the component perpendicular to the plane.
Introducing the angle $\theta(t)$ of the principal axis $a_1$ relative to $\vec{R}$, the relative
angular frequency $\Omega(t)$ of the $a_1$ axis to $\vec{R}$ is $\Omega = \frac{d\varphi}{dt} + \frac{d\theta}{dt}$.

Momentum conservation of the star leads to the following equations.
\begin{equation}
	\frac{d^2 a_1}{dt^2} = a_1(\Omega^2+\lambda^2) + 2a_2\Omega\lambda - 2a_1A_1 + 2\frac{p_c}{\rho a_1}
		+ \frac{GM_{\rm BH}a_1}{R^3}(3\cos^2\theta-1),
		\label{eq: eq for a1}
\end{equation}
\begin{equation}
	\frac{d^2 a_2}{dt^2} = a_2(\Omega^2+\lambda^2) + 2a_1\Omega\lambda - 2a_1A_2 + 2\frac{p_c}{\rho a_2}
		 + \frac{GM_{\rm BH}a_2}{R^3}(3\sin^2\theta-1),
	\label{eq: eq for a2}
\end{equation}
\begin{equation}
	\frac{d^2 a_3}{dt^2}= - 2a_3A_3 + 2\frac{p_c}{\rho a_3} - \frac{GM_{\rm BH}a_3}{R^3}.
	\label{eq: eq for a3}
\end{equation}

Conservation of angular momentum and vorticity is written as,
\begin{equation}
	\frac{d\Omega}{dt} = \frac{1}{a_1^2-a_2^2}
	\left[-2\left(a_1\frac{da_1}{dt}-a_2\frac{da_2}{dt}\right)\Omega\right.
	\left. - 2\left(a_1\frac{da_2}{dt}-a_2\frac{da_1}{dt}\right)\lambda\right.
	\left. -\frac{3GM_{\rm BH}\sin 2\theta}{2R^3}(a_1^2+a_2^2)
	\right]
	\label{eq: eq for Omega}
\end{equation}
and
\begin{equation}
	\frac{d\lambda}{dt} = \frac{1}{a_2^2-a_1^2}
	\left[-2\left(a_2\frac{da_1}{dt}-a_1\frac{da_2}{dt}\right)\Omega\right.
	\left. - 2\left(a_2\frac{da_2}{dt}-a_1\frac{da_1}{dt}\right)\lambda\right.
	\left. -\frac{3GM_{\rm BH}\sin 2\theta}{2R^3}a_1a_2
	\right]
	\label{eq: eq for omega}
\end{equation}
where $\rho(t)$ and $p_c(t)$ is the uniform density and the central pressure. $A_1$ to $A_3$ are defined as \citep{ChandraEFE}:
\begin{eqnarray}
	A_1 &=& \pi G\rho\frac{2a_1a_3}{a_1^2\sin^3\Phi\sin^2\Theta} [F(\Theta,\Phi)-E(\Theta,\Phi)], \\
	A_2 &=& \pi G\rho\frac{2a_2a_3}{a_1^2\sin^3\Phi\sin^2\Theta\cos^2\Theta}\left[E(\Theta,\Phi)-F(\Theta,\Phi)\right.
	\left.-\frac{a_3}{a_2}\sin^2\Theta\sin\Phi\right],\\
	A_3 &=& \pi G\rho \frac{2a_2a_3}{a_1^2\sin^3\Phi\cos^2\Theta}\left[\frac{a_2}{a_3}\sin\Phi - E(\Theta,\Phi)\right],
\end{eqnarray}
where $\sin\Theta \equiv \sqrt{(a_1^2-a_2^2)/(a_1^2-a_3^2)}$ and $\cos\Phi \equiv a_3/a_1$.  
$E(\Theta,\Phi)$ and $F(\Theta,\Phi)$ are incomplete elliptic integrals of the second and the first kind \citep{ChandraEFE}.

We assume the star is composed of ideal gas whose equation of state is written as
\begin{equation}
	p_c = \frac{k_{\rm B} \tau_c}{\mu m_{_{\rm H}}}\rho,
\end{equation}
where $k_{\rm B}, m_{_{\rm H}}, \mu$ are the Boltzmann's constant, the atomic mass unit and the mean molecular weight (to be fixed as $0.5$).
$\tau_c$ is the central temperature. It should be noticed that the profiles of pressure and temperature are quadratic in the
Cartesian coordinate \citep{Fujimoto1968ApJ...152..523F}.

We assume the fluid configuration changes adiabatically. Thus the first law of thermodynamics leads to
\begin{equation}
	\frac{1}{\tau_c}\frac{d\tau_c}{dt} = -\sum_{i=1}^3 \frac{1}{a_i}\frac{da_i}{dt}.
	\label{eq: eq for Tc}
\end{equation}

Eqs.(\ref{eq: eq for a1}),(\ref{eq: eq for a2}), (\ref{eq: eq for a3}), (\ref{eq: eq for Omega}), (\ref{eq: eq for omega}), (\ref{eq: eq for Tc})
are numerically integrated for given orbital parameters $R(t)$ and $\varphi(t)$. 
Here we choose a parabolic orbit around the black hole. Far from the periastron we put the initial star which is
an axisymmetric spheroid and whose spin axis is perpendicular to the orbital plane. 

After the periastron passage, we assume the star tidally disrupted\footnote{Notice that the tidal disruption event itself may not be described by
the simple ellipsoid model.}. We compute the total specific energy $E$ of fluid element of the star and compute its mass distribution $dM/dE$ .
The mass distribution translates to the mass infall rate $dM/dT$ by Eq.(\ref{eq: mass infall rate}).

\begin{figure}
\centering
\includegraphics{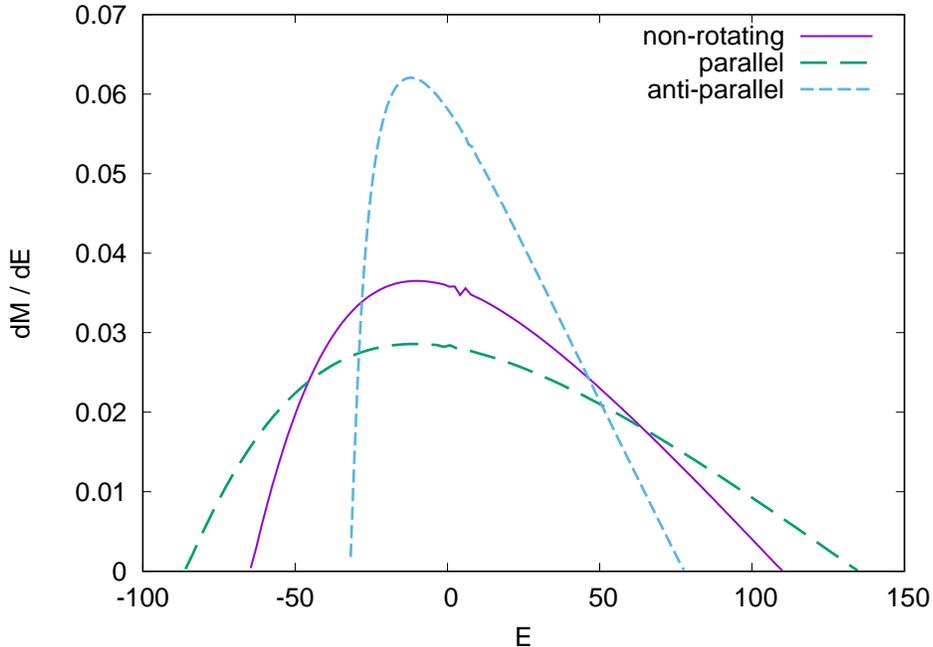}
\caption{Mass distribution of tidally-disrupted star as a function of specific energy. The specific energy $E$ is normalized by
$GM_\odot/R_\odot$, while the mass is normalized by $M_\odot$.}
\label{fig: ellipsoid dMdE 0.95}
\end{figure}

\begin{figure}
\centering
\includegraphics{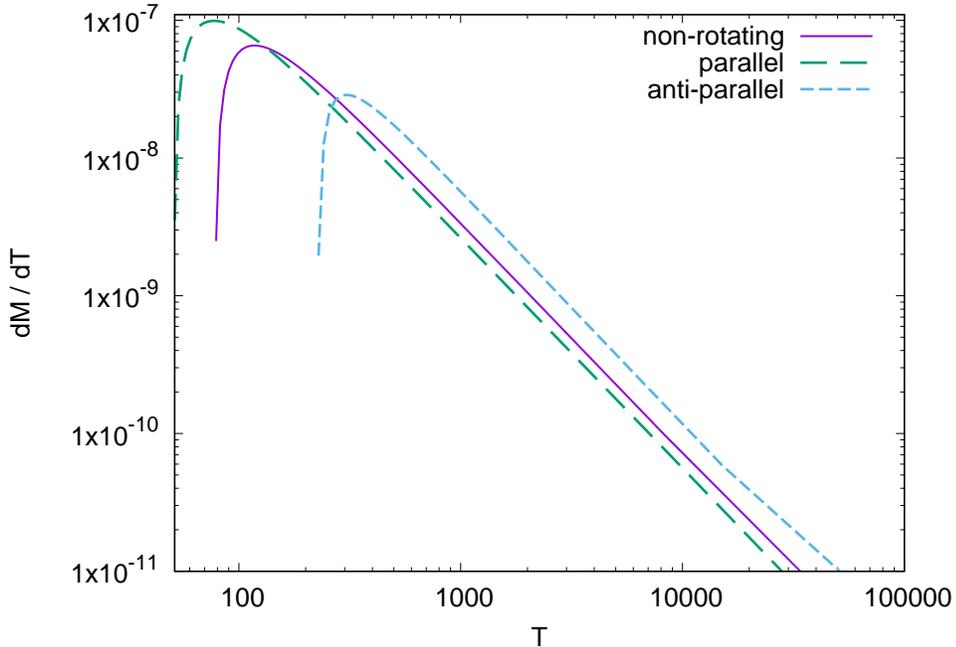}
\caption{Mass infall rate $dM/dT$ (Eq.(\ref{eq: mass infall rate})) is plotted as a function of infall
time $T$ (in units of day). The infall rate is in units of $M_\odot {\rm s}^{-1}$.}
\label{fig: ellipsoid dMdT 0.95}
\end{figure}

For an illustration,  we compare a zero spin star model of $m_\star=4M_\odot$ with a rapidly rotating
star of the same mass. The ratio of the semi-minor to the semi-major axis is $0.95$ for the rotating
star (the so-called $T/|W|$ parameter, the ratio of kinetic energy to gravitational energy, is $1.36\times 10^{-2}$.
Angular frequency of the model is $4.2\times 10^{-5}$Hz, which corresponds to the surface velocity of
$67~{\rm km}{\rm s}^{-1}$.). 
We fix the average radius $r_{\rm av} = (a_1a_2a_3)^{1/3}$ of the initial stars. The initial central
temperature of rotating star is $\tau_c=9.7\times 10^6$K, while that of the zero spin star
is $1.0\times 10^7$K. The periastron distance of the orbit is $2 (M_{\rm BH}/m_\star)^{1/3}R_\odot$
which corresponds to 87\% of the tidal radius of the star.
We compute $dM/dE$ and $dM/dT$ when the orbital phase is $\varphi = \pi/6$ after
the periastron. The cases with the spin axis parallel and anti-parallel to the orbital angular momentum
are compared (Fig.\ref{fig: ellipsoid dMdE 0.95}, Fig.\ref{fig: ellipsoid dMdT 0.95}).

%
%

Since the ratio of the tidal radius to the Schwarzschild radius scales as $M_{\rm BH}^{-2/3}$, the ratio
may be close to unity for $M_{\rm BH}=10^6M_\odot$. For the Newtonian models above, the ratio is 1/9. 
Thus the general relativistic effect may be important in the tidal disruption processes considered here. 
We assess the effect by introducing the modified gravitational potential (pseudo-Newtonian, 
or Paczy\'{n}ski-Wiita potential. See \cite{Paczynski_Wiita1980}) to mimic the strong gravity around 
the central black hole.
\begin{figure}
\centering
\includegraphics{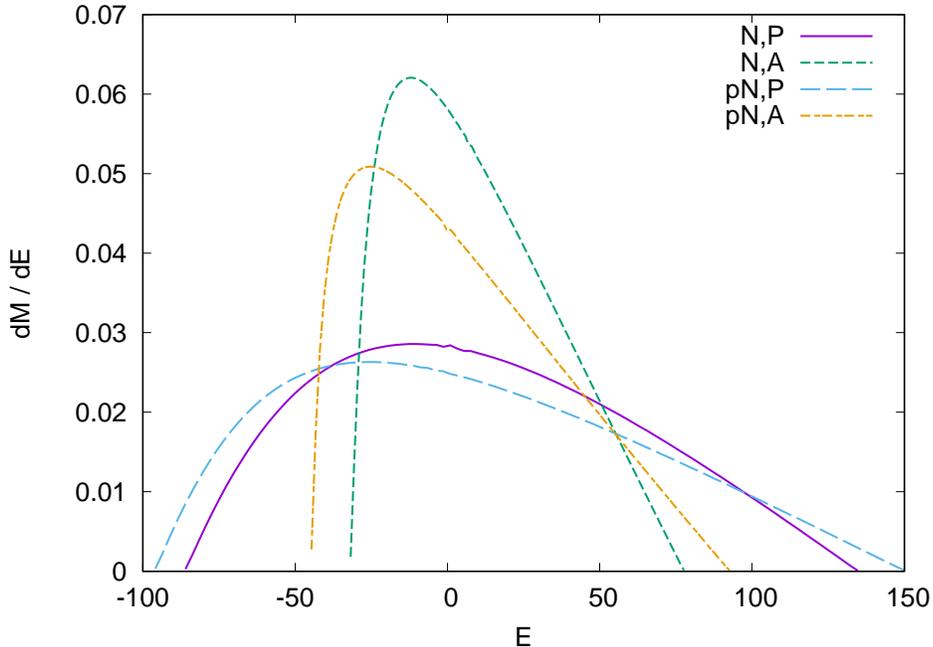}
\caption{Comparison of Newtonian and pseudo-Newtonian potential of the central black hole.
'N' stands for Newtonian, 'pN' for pseudo-Newtonian, 'P' for parallele spin, and 'A' for anti-parallel spin.}
\label{fig: dMdE 0.95 pseudo-Newton}
\end{figure}
\begin{figure}
\centering
\includegraphics{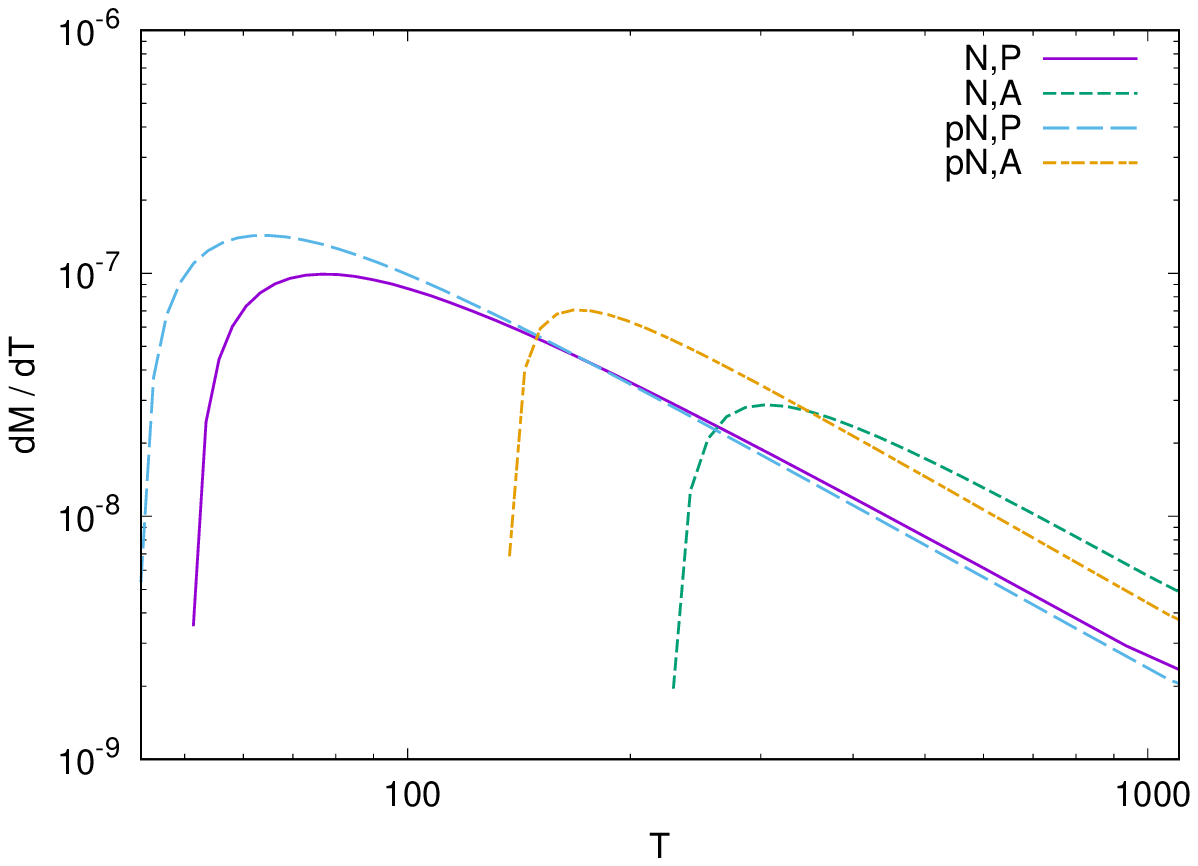}
\caption{Mass infall rate $dM/dT$ for the model in Fig.\ref{fig: dMdE 0.95 pseudo-Newton}.}
\label{fig: dMdT 0.95 pseudo-Newton}
\end{figure}
In Fig.\ref{fig: dMdE 0.95 pseudo-Newton} we compare the Newtonian (N) and pseudo-Newtonian (pN)
cases for parallel (P) and anti-parallel (A) spins. The axis ratio of the star is $0.95$ and the periastron
distance is $12$ times the Schwarzschild radius. In each case of parallel or anti-parallel spin, the mass
spreads in wider energy range for the pseudo-Newtonian model than for the corresponding Newtonian
case. As a result, the mass infall rate rises
earlier for the pseudo-Newtonian potential (Fig.\ref{fig: dMdT 0.95 pseudo-Newton}). It reflects the fact
that the gravity and the tidal force of the pseudo-Newtonian case are stronger than those of the Newtonian
case. The tidally-disrupted fluid elements bound to the black hole fall deeper in the potential well and
result in the earlier rise of the mass infall rate.  

\begin{figure}
\centering
\includegraphics{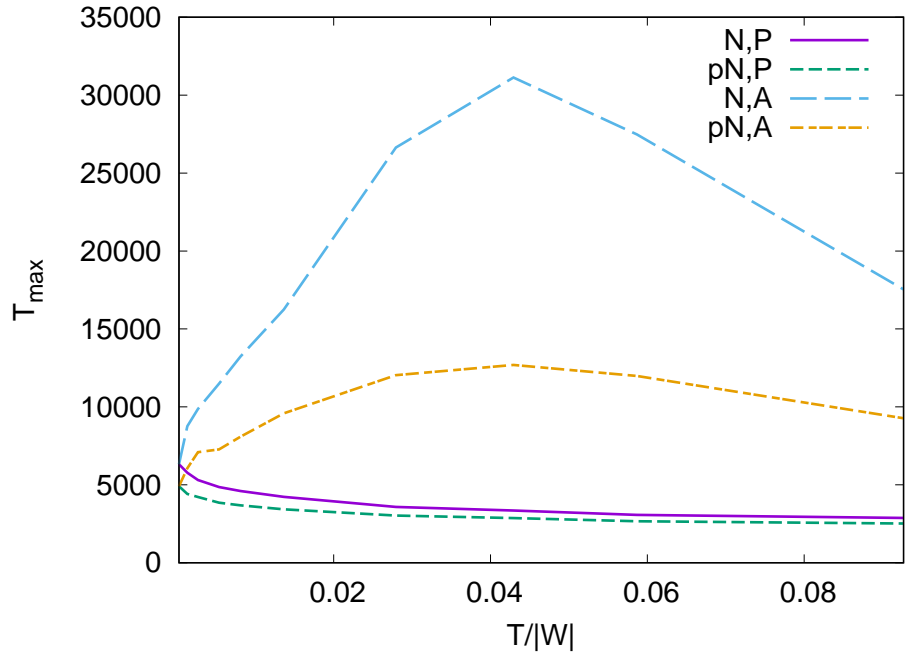}
\caption{The elapsed time between the disruption and the maximum of mass infall rate, 
$T_{\rm max}$ (in units of day), as a function of dimensionless parameter $T/W$.
'N' is for Newtonian potential, while 'pN' is for pseudo-Newtonian
potential for the central black hole. 'P' and 'A' are parallel and anti-parallel spin cases.}
\label{fig: Tmax}
\end{figure}

In Fig.\ref{fig: Tmax} we plot as a function of dimensionless parameter $T/W$ the time $T_{\rm max}$
between the periastron passage and the maximum of mass infall rate. We see that neglecting the strong gravity
of the central black hole overestimates $T_{\rm max}$. It is reasonable since the strong gravity tends to
bind more mass to the central objects. However, rotation of a star is also not to be neglected
if $T/W$ parameter is$\sim{\cal O}[10^{-2}]$. The difference of $T_{\rm max}$
between parallel and anti-parallel spins is smaller for stronger gravity of pseudo-Newtonian potential.

\begin{acknowledgments}
We thank Takeru Suzuki, Izumi Hachisu, and Nicholas Stone for useful discussions and
comments.  SY was supported by JSPS Grant-in-Aid for Scientific
Research(C) 18K03641. AT was supported by JSPS Grant-in-Aid for Young
Scientists (B) (16K17656), and by JSPS Grant-in-Aid for Innovative
area (17H06360). Numerical analyses were in part carried out on computers 
at Center for Computational Astrophysics, National Astronomical Observatory of Japan.
\end{acknowledgments}


\bibliography{Kagaya}

\end{document}